\documentclass[aps,prd,reprint,nofootinbib,amsmath]{revtex4-1}
\usepackage{amsmath,amssymb,amsthm,amsxtra,overpic,bm,epsfig,subfigure}
\usepackage{mathrsfs}
\usepackage{graphicx}
\usepackage{color}
\usepackage{comment}
\usepackage{epstopdf}
\usepackage{enumitem}
\usepackage{float}
\usepackage[colorlinks,
linkcolor=blue, urlcolor=blue, anchorcolor=blue, citecolor=blue]{hyperref}

\newcommand{\PLH}{{\mkern-2mu\times\mkern-2mu}}

\begin{document}

\title{Neutrinophilic Axion-Like Dark Matter}
\author{Guo-yuan Huang}
\email{huanggy@ihep.ac.cn} 
\author{Newton Nath}
\email{newton@ihep.ac.cn}
\affiliation{Institute of High Energy Physics, and School of Physical
Sciences}
\affiliation{University of Chinese Academy of Sciences, Beijing 100049, China}

\begin{abstract}
 The axion-like particles (ALPs) are very good candidates of the cosmological dark matter, which can exist in many extensions of the standard model (SM). The mass range of the ALPs as the dark matter can extend from a sub-eV scale to almost $10^{-22}~{\rm eV}$. On the other hand, the neutrinos are found to be massive and the SM must be extended to explain the sub-eV neutrino masses. {It becomes very interesting to consider an exclusive coupling between these two low scale frontiers that are both beyond the SM.}
 The propagation of neutrinos inside the Milky Way would undergo the coherent forward scattering effect with the ALP background, and the neutrino oscillation behavior can be modified by the ALP-induced potential. Assuming a derivative coupling between the ALP and the three generations of active neutrinos, possible impacts on the neutrino oscillation experiments have been explored in this paper. In particular, we have numerically studied the sensitivity of the Deep Underground Neutrino Experiment (DUNE). The astrophysical consequences of such coupling have also been investigated systematically.
\end{abstract}

\maketitle
\section{Introduction}
As one of the most promising dark matter candidates, the ALP with a possible mass spanning from sub-eV to $\mathcal{O}(10^{-22})~{\rm eV}$ \footnote{The ultralight dark matter with a mass $\sim 10^{-22}~{\rm eV}$ is often called ``fuzzy dark matter'' \cite{Hu:2000ke}. Its macroscopic wavelength can suppress the power of the galactic clustering and resolve the small scale tension of the cold dark matter paradigm \cite{Hui:2016ltb}. A lower bound on the ALP mass can be set by the observation of the Lyman-$\alpha$ forest at $\mathcal{O}(10^{-22})~{\rm eV}$ \cite{Irsic:2017yje,Desjacques:2017fmf}.} is drawing more and more attention (see \cite{Kim:2008hd,Graham:2015ouw,Irastorza:2018dyq,Marsh:2015xka} for recent reviews). The typical QCD axion \cite{Weinberg:1977ma,Wilczek:1977pj} was first identified as the pseudo-Nambu-Goldstone (pNG) boson in the Peccei-Quinn (PQ) mechanism \cite{Peccei:1977hh,Peccei:1977ur} to solve the strong CP problem of QCD. It was later realized \cite{Dine:1982ah,Abbott:1982af,Preskill:1982cy,Davis:1986xc,Kim:1986ax} to be a very good candidate of the cold dark matter which can acquire an effective anomalous mass by interacting with the gluons. The mass of the QCD axion is settled by the QCD phase transition scale $\Lambda^{}_{\rm QCD}$ and the PQ scale $f_a$, e.g. $m_a \approx \Lambda^2_{\rm QCD}/f_a \approx 10^{-5}~{\rm eV}~(10^{12}~{\rm GeV}/f_a)$ with $\Lambda^{}_{\rm QCD} \approx \mathcal{O}(10^2)~{\rm MeV}$ and $f_a \approx 10^{9}$-$10^{12}~{\rm GeV}$. Moreover, the axion can interact with the standard model (SM) fermions through a derivative type of coupling whose strength is proportional to its mass $m^{}_{a}$.
On the other hand, the ALPs with very similar properties to the QCD axion are well motivated in many other extensions of the SM. For example, generating the tiny neutrino mass by spontaneously breaking the lepton number \cite{Chikashige:1980qk,Chikashige:1980ui,Gelmini:1980re,Choi:1991aa,Acker:1992eh,Georgi:1981pg,Schechter:1981cv} predicts the existence of a Nambu-Goldstone (NG) boson, the Majoron, which inevitably couples with the neutrinos. To spontaneously break the family symmetries will lead to the familons \cite{Wilczek:1982rv,Reiss:1982sq}. In the string theory framework, many different ALPs can appear naturally \cite{Witten:1984dg,Choi:1985je,Svrcek:2006yi,Conlon:2006tq,Arvanitaki:2009fg,Cicoli:2012sz}, and one of them could just be the QCD axion. In some unified models \cite{Dias:2014osa,Ballesteros:2016xej,Ema:2016ops,Calibbi:2016hwq,Bjorkeroth:2017tsz,Bjorkeroth:2018dzu,Ema:2018abj}, the pseudoscalar particle can even play multiple roles among the QCD axion, the majoron and the familon at the same time. 

The mass spectrum and the coupling strength with the SM particles of the generic ALPs are not limited as in the QCD axion model, greatly enriching their phenomenology and the experimental efforts to hunt them. The ALPs can be non-thermally produced in the early Universe by the misalignment mechanism or the topological defect decay. After the ALPs form the dark matter halo of our Milky Way, their de Broglie wavelength $\lambda_{\rm dB}$ can be as long as several kpc for an ALP mass $m_{a} \approx 10^{-22}~{\rm eV}$, almost comparable to the scale of the Milky Way $\sim \mathcal{O}(10)$~{\rm kpc}. The local number density of the ultralight particles is $n_a = \rho_a / m_a \approx 10^{30}~{\rm cm^{-3}}$ if they saturate the dark matter energy density $\rho \approx 0.3~{\rm GeV \cdot cm^{-3}}$. The huge particle number within the de Broglie wavelength makes them oscillate coherently as a single classical field with
 \begin{eqnarray}
 \label{eq:atx}
 a(t,x) = a_0 \cos{(m_a t - \vec{p} \cdot \vec{x})}, \;
 \end{eqnarray}
  where $a_0 = \sqrt{2\rho}/m_a$ is the amplitude of the ALP field, $\vec{p}$ is the momentum of the ALP, and $m_a$ denotes the mass of the ALP. The ALP field maintains its coherence during the expansion of the universe.
However, after the formation of the Milky Way, it develops a slight stochastic decoherence due to the gravitational clustering effect. The typical velocity dispersion of the ALP dark matter is $\sigma_{v} \sim 10^{-3} c$ \cite{Lentz:2017aay}, corresponding to a coherence length of $\lambda^{}_{\rm coh}\approx \lambda^{}_a \times 10^3$ with $\lambda^{}_{a} = 2\pi / m^{}_a$ being its Compton wavelength. In our work, we consider an ALP-neutrino interacting Lagrangian which preserves the shift symmetry as 
\begin{eqnarray}
\label{eq:L}
-\mathcal{L}_{\rm int} = g^{}_{\alpha \beta} \partial_{\mu}a~ \overline{\nu}_{\alpha}\gamma^{\mu}\gamma_5 \nu_{\beta}, \;
\end{eqnarray}
where $g^{}_{\alpha \beta}$ with $ \alpha, \beta = (e, \mu, \tau) $ is the dimensional coupling strength in the neutrino flavor basis with $g^{}_{\alpha \beta} = g^{*}_{\beta\alpha}$, and $\nu_{\alpha(\beta)}$ represents the neutrino flavor eigenstate. In the QCD axion model, the coupling strength $g^{}_{\alpha \beta}$ is suppressed by the PQ scale $f^{}_{a}$, which is not necessarily the case in the generic ALPs context.

The neutrino propagating in the Milky Way can coherently interact with the neutrinophilic ALP field, and the mixings among the neutrino flavors will be modified effectively. In this note, we will study systematically the impacts of the ALP-neutrino coupling on the neutrino oscillation experiments as well as the astrophysical phenomenology. In particular, we will take DUNE as a typical example, and numerically study its sensitivity. The influences of the possible ultralight dark matter field on the neutrino phenomenology have been discussed from various perspectives in the literature   \cite{Berlin:2016woy,Brdar:2017kbt,Krnjaic:2017zlz,Liao:2018byh,Capozzi:2018bps,Davoudiasl:2018hjw,Reynoso:2016hjr}. 
However, it is still very worthwhile to conduct a work like this one. In the existing works, the ultralight scalar        \cite{Berlin:2016woy,Brdar:2017kbt,Krnjaic:2017zlz,Liao:2018byh,Capozzi:2018bps,Davoudiasl:2018hjw,Reynoso:2016hjr}, vector \cite{Brdar:2017kbt,Capozzi:2018bps} and tensor \cite{Capozzi:2018bps} DMs have been considered for the neutrino oscillation experiments. The pseudoscalar ALP with a derivative coupling is now being investigated in this work. We will see the derivative coupling in Eq.~(\ref{eq:L}) can have very different laboratory and astrophysical consequences.  
 The impacts on the neutrino oscillation experiments greatly depend on the duration of one ALP oscillation cycle. The time-dependent perturbation analysis should be adopted when the ultralight DM oscillates a number of cycles during the neutrino propagation. If the ALP field loses its coherence within a single neutrino flight, the forward scattering effect which is coherently enhanced should vanish stochastically. 
  Various astrophysical constraints on the coupling are considered systematically in this work. The free-streaming constraint of the cosmic microwave background (CMB) for the neutrino decay process should be the most stringent one. The influences on the propagation of supernova neutrinos and ultrahigh-energy (UHE) neutrinos are also discussed in detail. 

This paper is organized as follows. In Section II, the influence of the ALP dark matter on the neutrino oscillation behavior have been explored in detail, and the impact on DUNE has been elaborated with a numerical sensitivity study. In Section III, we will generally discuss the existing constraints from astrophysical observations. We make our conclusion in Section IV. 

\section{Impacts on Neutrino Oscillations}
When neutrinos propagate inside our galaxy, which is always true for the current ground-based oscillation experiment, they will inevitably scatter with the ambient ALP background under our consideration. The coherent forward scattering process which can be analogous to the Mikhyev-Smirnov-Wofenstein (MSW) \cite{Wolfenstein:1977ue,Mikheev:1986gs,Mikheev:1986wj} matter effect is dominant. Under the interaction form in Eq.~(\ref{eq:L}), the dispersion relation of the three generations of neutrinos will be modified as
\begin{eqnarray}
\label{eq:dispersion}
(E_{\nu} \cdot \hat{\rm I}+  \partial_0 a \cdot {g^{}_{\rm m}})^2 - ( \vec{p}_{\nu} \cdot \hat{\rm I} + \vec{\partial}a \cdot {g^{}_{\rm m}})^2 = {m}^2_{\nu},
\end{eqnarray}
where $\hat{\rm I}$ is the $3\PLH 3$ identity matrix, ${g^{}_{\rm m}}$ stands for the $3 \PLH 3$ Hermitian matrix of the coupling constant $g_{\alpha\beta}$, ${m}^{}_{\nu}$ stands for the mass matrix of neutrinos in the flavor basis, $E^{}_{\nu}$ and $\vec{p}^{}_{\nu}$ are the energy and momentum of the neutrino, respectively. The modified neutrino oscillation behavior can be conveniently described by the following effective Hamiltonian:
\begin{eqnarray} \label{eq:Hm}
H(t) & = & \frac{1}{2E^{}_{\nu}} U \left(\begin{matrix} 0 & 0 & 0 \\ 0 & \Delta m^2_{21} & 0 \\ 0 & 0 & \Delta m^2_{31} \end{matrix}\right)  U^\dagger_{} \\ \nonumber 
& &+\left(\begin{matrix} V+\xi^{}_{ee}(t) & \xi^{}_{e\mu}(t)& \xi^{}_{e\tau}(t) \\ \xi^{*}_{e\mu}(t) & \xi^{}_{\mu\mu}(t) & \xi^{}_{\mu\tau}(t) \\ \xi^{*}_{e\tau}(t) & \xi^{*}_{\mu\tau}(t) & \xi^{}_{\tau\tau}(t) \end{matrix}\right) \; ,
\end{eqnarray}
where $U$ stands for the flavor mixing matrix of the three generations of neutrinos in vacuum, $\Delta m^{2}_{21}$ and $\Delta m^{2}_{31}$ are the neutrino mass-squared differences in vacuum. Here $V$ is a potential characterizing the MSW effect with the possible ordinary matter around the neutrino, and $\xi^{}_{\alpha\beta}$ for $\alpha,\beta=(e,\mu,\tau)$ is the potential contributed by the ALP background. $\xi^{}_{\alpha\beta}$ can be derived from Eq.~(\ref{eq:dispersion}), which at the leading-order reads
\begin{eqnarray}
\label{eq:Veff}
\xi^{}_{\alpha\beta} \approx -{g}^{}_{\alpha\beta}\partial_0 a + {g}^{}_{\alpha\beta} \vec{\partial}a \cdot \vec{p}^{}_{\nu}/{|\vec{p}^{}_{\nu}|}, \;
\end{eqnarray}
where $\vec{p}^{}_{\nu}/{|\vec{p}^{}_{\nu}|}$ stands for the direction of the neutrino flux. Providing the velocity of the ALP field in the Milky Way is of the order $\mathcal{O}(10^{-3})c$, the temporal term should dominate the potential in Eq.~(\ref{eq:Veff}). The further approximation can be made to give 
\begin{eqnarray}
\label{eq:Veffapp}
\xi^{}_{\alpha\beta} (t) \approx - g^{}_{\alpha\beta}\partial_0 a \approx g^{}_{\alpha\beta}\sqrt{2\rho} \sin{m^{}_{a} t}, \;
\end{eqnarray}
with $\sqrt{2\rho} \approx 2.15\times 10^{-3}~{\rm eV^2}~\sqrt{\rho/(0.3~{\rm GeV\cdot cm^{-3}})} $. As has been mentioned before, the ALP potential oscillates with a period $t^{}_{a} \approx 2\pi/m^{}_{a} \approx 4.8~{\rm day}~(10^{-20}~{\rm eV}/m^{}_{a})$. The impact on the neutrino oscillation strongly depends on the magnitude relation between the ALP period and several time scales of the experiment. The relevant experimental time scales include: $  t^{}_{\rm opr}$, the operation time of the experiment, typically over several years for an oscillation experiment; $t^{}_{\rm rsv}$, the minimal period that the experiment can resolve for a periodic modulation effect; $t^{}_{\rm flt}$, the flight time of each single neutrino from the source to the detector, $\sim 0.33~{\rm ms}$ for a baseline length of $100~{\rm km}$. There are four different cases that should be addressed:
\begin{itemize}[noitemsep,topsep=0pt,leftmargin=5.5mm]
\item[(i)] {Case I}: $t^{}_{\rm opr} \ll t^{}_{a}$. In this case, neutrinos will experience an approximately constant ALP-induced potential throughout the operation time of the experiment. The influence to neutrino oscillations is very similar to the non-standard interactions (NSIs) of neutrinos \footnote{Wolfenstein in Ref.~\cite{Wolfenstein:1977ue} first proposed that dimension-six four-fermion operators in the form of NSI can potentially impact the neutrino propagation.} with a constant matter density profile. But the effect of the ALP field is irreducible for all oscillation experiments in our galaxy, even with no ordinary matter surrounding the neutrino. For an ALP mass $m^{}_{a} \approx 10^{-22}~{\rm eV}$ which might be the minimal feasible mass to form the dark matter, the corresponding period is $t^{}_{a} \approx 1.3~{\rm yr}$. However, the operation time of most oscillation experiments is longer than $1.3~{\rm yr}$. For them, the ALP-induced potential is no longer constant during the operation.
\item[(ii)] {Case II}: $t^{}_{\rm rsv} \lesssim t^{}_{a} \lesssim t^{}_{\rm opr}$. The operation time is longer than the period of the ALP field, so it can cover a number of ALP cycles during the operation. To detect the modulation effect of the oscillating potential, the duration of each ALP cycle must be longer than the experimental resolution limit of the periodicity. There are several experiments that have looked for the periodic modulation effect of the solar neutrinos, e.g. Super-Kamiokande \cite{Yoo:2003rc,Sturrock:2005wf,Ranucci:2005ep,Ranucci:2007fb} and SNO \cite{Ranucci:2007fb,Ranucci:2006rz,Aharmim:2005iu,Collaboration:2009qz}. The period of possible periodic signals have been scanned from $\sim 10~{\rm min}$ to $\sim 10~{\rm yr}$. No anomalous modulation effect is found in the data sets of Super-Kamiokande and SNO using various statistical methods. Based on this, Ref.~\cite{Berlin:2016woy} has set a strong constraint on the ultralight scalar coupling. The minimal recognizable period $t^{}_{\rm rsv}$ of an experiment depends on many factors: the time binning of the data, the events number, the statistical approach etc.. In principle, $t^{}_{\rm rsv}$ can be as small as the precision of the time measurement, e.g. $\sim 100~{\rm ns}$ for SNO \cite{Collaboration:2009qz}. The statistical analysis in the solar case can be similarly applied to other types of oscillation experiments, to locate the ALP-induced signal. Moreover, in this case one can also choose to integrate over the data-taking time to get a smaller averaged effect \cite{Brdar:2017kbt,Krnjaic:2017zlz,Liao:2018byh}.
\item[(iii)] {Case III}: $t^{}_{\rm flt} \ll t^{}_{a} \lesssim t^{}_{\rm rsv}$. The ALP field oscillates so fast that the experiment can no longer resolve the periodic modulation effect. But during each single neutrino flight from source to detector, the ALP field is still approximately constant. There is no other way but to integrate over the data-taking time and obtain the averaged distortion effect.
\item[(iv)] {Case IV}: $t^{}_{\rm a} \lesssim t^{}_{\rm flt} \lesssim t^{\prime}_{\rm a}$. Throughout a single neutrino flight, the ALP field has been oscillating for a number of cycles. In this case, the neutrino flavor evolution is driven by a varying potential. The time-dependent perturbation theory is required for the perturbative expansion analysis. However, if the neutrino flight time is even longer than the decoherence time of the ALP $t^{\prime}_{\rm a} \approx 1000 \times t^{}_{\rm a}$, i.e. $t^{}_{\rm flt} > t^{\prime}_{a}$, the coherent potential induced by the ALP field should vanish stochastically.
\end{itemize}

\subsection{Perturbative Expansion}
We will work under the two-neutrino-flavor scheme to see the influence of the ALP potential analytically. Let us assume the ALP-induced potential is very small, such that one can perturbatively expand the effective mixing parameters and the oscillation probability around the standard ones. Taking $\nu^{}_{e}$ and $\nu^{}_{\mu}$ as the two neutrino flavors under our consideration, the effective Hamiltonian in Eq.~(\ref{eq:Hm}) is then reduced to \footnote{The ordinary matter effect is temporarily ignored for the perturbation analysis. It is very straightforward to include it by replacing the vacuum quantities with the matter-corrected ones.}
\begin{eqnarray} \label{eq:Hm2f}
& &\frac{1}{2E^{}_{\nu}} \tilde{U}(t) \left(\begin{matrix} 0 & 0 \\ 0 & \Delta \tilde{m}^2(t) \end{matrix}\right)  \tilde{U}^\dagger_{}(t) = \\ \nonumber
& & \hspace{5 mm} \frac{1}{2E^{}_{\nu}} \left[U \left(\begin{matrix} 0 & 0 \\ 0 & \Delta m^2 \end{matrix}\right)  U^\dagger_{}+2E^{}_{\nu}\left(\begin{matrix} \xi^{}_{ee}(t) & \xi^{}_{e \mu}(t) \\ \xi^{*}_{e \mu}(t) &  \xi^{}_{\mu\mu}(t) \end{matrix}\right) \right],
\;
\end{eqnarray}
where $\Delta \tilde{m}^2_{}(t)$ is the effective mass-squared difference which has been shifted by the ALP background, and $\tilde{U}(t)$ is the effective mixing matrix that diagonalizes the total Hamiltonian.
Setting $\theta$ and $\tilde{\theta}$ as the mixing angles of the matrices ${U}$ and $\tilde{U}(t)$ respectively, one can formally expand the effective quantities to the second-order as
\begin{eqnarray} \label{eq:dtm1}
\Delta \tilde{m}^2_{} &=& \Delta m^2_{} + \delta^{}_{1}(\Delta m^2_{})+\delta^{}_{2}(\Delta m^2_{}), \\ \nonumber
\tilde{\theta} &=&\theta +\delta^{}_{1}(\theta) + \delta^{}_{2}(\theta).
\end{eqnarray}
Here $\delta^{}_{1}$ and $\delta^{}_{2}$ represent the first-order and the second-order ALP perturbations respectively, which read as follows
\begin{eqnarray} \label{eq:delta}
{\delta^{}_{1}(\Delta m^2)} &=& 2\left(\overline{\xi}\cos{2\theta}+2 \xi^{\rm R}_{e\mu} \sin{2\theta} \right){E^{}_{\nu}}, \\
\delta^{}_{1}( \theta) &=& \frac{2 \xi^{\rm R}_{e\mu} \cos{2\theta} -\overline{\xi} \sin{2\theta}}{\Delta m^2} {E^{}_{\nu}} , \\
\delta^{}_{2}(\Delta m^2)&=&
\biggl[ 4 (\xi^{\rm I}_{e\mu})^2 +4 (\xi^{\rm R}_{e\mu})^2 \cos^2{2\theta}  \\ \nonumber
& & \hspace{5 mm}  + \overline{\xi}^2\sin^2{2\theta} - 2 \xi^{\rm R}_{e\mu} \overline{\xi} \sin{4\theta} \biggr] \frac{2E^{2}_{\nu}}{ \Delta m^2}, \\
\delta^{}_{2}(\theta) &=&  \biggl[ -4 \xi^{\rm R}_{e\mu} \overline{\xi} \cos{4\theta} + 4 (\xi^{\rm I}_{e\mu})^2 \cot{2\theta} \\ \nonumber
& & \hspace{5 mm} + \overline{\xi}^2\sin{4\theta} - 4 (\xi^{\rm R}_{e\mu})^2 \sin{4\theta} \biggr] \frac{{E^{2}_{\nu}}}{ \left({\Delta m^2}\right)^2} , 
\end{eqnarray}
with $\overline{\xi} \equiv \xi^{}_{\mu\mu} - \xi^{}_{ee}$, and $\xi^{\rm R (I)}_{e\mu}$ being the real (imaginary) part of $\xi^{}_{e\mu}$. One can observe that $\Delta m^2$ and $\theta$ have been shifted by amounts of $\xi E^{}_{\nu}$ and $\xi E^{}_{\nu}/ \Delta m^2$ respectively, with $\xi$ denoting the general order of magnitude of $\xi^{}_{\alpha\beta}$.
Assuming the ALP potential is constant within $t^{}_{\rm flt}$ (for {Case I}, {Case II} or {Case III}), the survival probability of $\nu^{}_{\mu} \rightarrow \nu^{}_{e}$ reads
\begin{eqnarray} \label{eq:pee}
\tilde{P}^{}_{\mu e} &=& P^{}_{\mu e}+\delta^{}_{1}(P^{}_{\mu e})+\delta^{}_{2}(P^{}_{\mu e}), \\ \nonumber
P^{}_{\mu e} &=&\sin^2{\phi} \sin^2{2{\theta}} ,
\end{eqnarray}
where $\phi \equiv {\Delta {m}^2 L}/(4E)$ is a dimensionless phase factor and the corrections are
\begin{eqnarray} \label{eq:peecr}
\delta^{}_{1}(P^{}_{\mu e}) &=&  \phi \sin{2\phi} \sin^2{2\theta} \times \frac{\delta^{}_{1} (\Delta m^2)}{\Delta m^2}  \\ \nonumber
& & +2 \sin^2{\phi} \sin{4\theta}  \times \delta^{}_{1}(\theta), \\ \nonumber
 \delta^{}_{2}(P^{}_{\mu e}) &=&\left(\phi\right)^2 \cos{2\phi} \sin^2{2\theta}  \times \left[\frac{\delta^{}_{1}(\Delta m^2)}{\Delta m^2}\right]^2  \\ \label{eq:peecr1}
 & & + 4 \sin^2{\phi}\cos{4\theta}  \times \left[ \delta^{}_{1}(\theta) \right]^2   \\ \nonumber
 & & + 2\phi \sin{2\phi}\sin{4\theta}  \times \frac{\delta^{}_{1} (\Delta m^2) \delta^{}_{1}(\theta)}{\Delta m^2}  \\ \nonumber
 & & + \phi  \sin{2\phi}\sin^2{2\theta}\times \frac{\delta^{}_{2}(\Delta m^2)}{\Delta m^2}  \\ \nonumber 
 & &  +2 \sin^2{\phi}\sin{4\theta} \times \delta^{}_{2}(\theta).
\end{eqnarray}
 When the ALP field oscillates fast enough like in {Case II} and {Case III},
 the first-order term $\delta^{}_{1}(P^{}_{\mu e}) \propto \sin{m^{}_{a}t}$ can be averaged to $0$. However, the second-order term after being averaged, $\delta^{}_{2}(P^{}_{\mu e}) \propto\left<\sin^2{m^{}_{a}t}\right> = 1/2$, will lead to the distortion effect on the final probability.
 It should be emphasized that the second-order corrections in Eq.~(\ref{eq:dtm1}) are as important as the first-order ones when we derive $\delta^{}_{2}(P^{}_{\mu e})$ in Eq.~(\ref{eq:peecr1}). We notice that $\delta^{}_{1}(P^{}_{\mu e})/P^{}_{\mu e} \approx \xi E^{}_{\nu}/\Delta m^2$ and $\delta^{}_{2}(P^{}_{\mu e})/P^{}_{\mu e} \approx (\xi E^{}_{\nu}/\Delta m^2)^2$ assuming $\phi,\theta \approx \mathcal{O}(1)$.
 
The above discussion is under the assumption that the ALP field is approximately constant during a single neutrino flight. For {Case IV}, the above results do not hold any more. In the case of the time-dependent potential, the transition amplitude $\mathcal{M} \equiv \langle \nu^{}_{e} | \nu^{}_{\mu} \rangle^{}_{t}$ can be formally expanded as $\mathcal{M} = \mathcal{M}_{0}+\mathcal{M}^{}_{1}+\mathcal{M}^{}_{2}+\cdot \cdot \cdot$ with $\mathcal{M}^{}_{i}$ being the $i$th-order result. Using the time-dependent perturbation theory, one can work out the following first-order amplitude correction,
\small
\begin{eqnarray} \label{eq:amp}
\mathcal{M}^{}_{1} &=& i\sum_{i j} \Lambda^{}_{ i j } \exp{\left(-i \hat{H}^{}_{0,ii} t^{}_{2}+i \hat{H}^{}_{0,jj} t^{}_{1}\right)} \\ \nonumber
& &\times \Biggl[ \frac{\exp{\bigl[i (\hat{H}^{}_{0,ii} -\hat{H}^{}_{0,jj}+m^{}_{a}) t \bigr] }}{2 (\hat{H}^{}_{0,ii} -\hat{H}^{}_{0,jj}+m^{}_{a})}  \\ \nonumber
& &  -\frac{\exp{\bigl[i (\hat{H}^{}_{0,ii} -\hat{H}^{}_{0,jj} - m^{}_{a}) t \bigr]}}{2 (\hat{H}^{}_{0,ii} -\hat{H}^{}_{0,jj} - m^{}_{a})}\Biggr]^{t=t^{}_{2}}_{t=t^{}_1},
\end{eqnarray}
\normalsize
where the subscripts $i,j$ run over $(1,2)$. Here $t^{}_{1}$ and $t^{}_{2}$ denote the initial time and final time of the neutrino flight, respectively, $\hat{H}^{}_{0,11} = 0$ and $\hat{H}^{}_{0,22} = \Delta m^2_{}/(2E^{}_{\nu})$ are the eigenvalues for the Hamiltonian in vacuum $H^{}_{0}$, and we define $\Lambda^{}_{i j} \equiv \sum_{\eta \gamma} U^{}_{e i} U^{\dagger}_{i\eta}  U^{}_{\gamma j} U^{\dagger}_{j\mu} g^{}_{\eta\gamma}\sqrt{2\rho} \approx \mathcal{O}(\xi)$ with $\eta,\gamma$ run over $(e,\mu)$. One can notice a singularity at the point $m^{}_{a} = \Delta m^2_{}/(2E^{}_{\nu})$ in the denominator. However, the singularity can be cancelled by the numerator and reduce to a factor $\sim t^{}_{\rm flt}$. The baseline of an oscillation experiment is usually selected around the oscillation maximum, i.e. $t^{}_{\rm flt} \equiv (t^{}_{2} - t^{}_{1}) \approx 2\pi E^{}_{\nu}/\Delta m^2_{}$. For Case IV with $t^{}_{a} \equiv 2\pi/m^{}_{a} < t^{}_{\rm flt}$, we should have $ \Delta m^2_{}/E^{}_{\nu} \lesssim m^{}_{a}$. One might as well further set $\Delta m^2_{}/(2E^{}_{\nu}m^{}_{a}) \lesssim \mathcal{O}(0.1)$, such that Eq.~(\ref{eq:amp}) can be approximated as
\small
\begin{eqnarray} \label{eq:ampSim}
\mathcal{M}^{}_{1} &\approx& i\sum_{i j} \frac{\Lambda^{}_{\alpha i j \beta}}{m^{}_{a}} \exp\left(-i \hat{H}^{}_{0,ii} t^{}_{2}+i \hat{H}^{}_{0,jj} t^{}_{1} \right)   \\ \nonumber
& &  \times \left[\cos{(m^{}_{a} t^{}_{2})} - \cos{(m^{}_{a} t^{}_{1})}\right].
\end{eqnarray}
\normalsize
The final survival probability of $\nu^{}_{\mu} \rightarrow \nu^{}_{e}$ is given by ${P}^{}_{\mu e} \approx |\mathcal{M}^{}_{0}|^2+2{\rm Re}(\mathcal{M}^{}_{o}\mathcal{M}^{*}_{1})+|\mathcal{M}^{}_{1}|^2$. As has been mentioned before, the flight time for a baseline length of $L=100~{\rm km}$ is $t^{}_{\rm flt} \approx 0.3~{\rm ms}$. For Case IV under the discussion, the oscillating period of ALP should be in the range of $\left(0.3~{\rm \mu s},0.3~{\rm ms}\right)$. If the experiment can resolve the periodic fluctuation in this time range (that would require a very refined scanning in the experimental data set), the modulation pattern with an amplitude $\xi/m^{}_{a}$ can be identified. Regardless of the practical periodicity resolving power, one can always choose to average over the observation time, obtaining the probability correction at the order of $\mathcal{O}(\xi/m^{}_{a})^2$. Note that the cross term ${\rm Re}(\mathcal{M}^{}_{o}\mathcal{M}^{*}_{1})$ vanishes for the averaged probability. 
 
 In conclusion, the ALP field will induce a potential $\xi \approx g \sqrt{2\rho} \sin{m^{}_{a} t}$ in the neutrino effective Hamiltonian, with $g$ denoting the general order of magnitude of the coupling strength $g^{}_{\alpha\beta}$. If the ALP field is approximately constant for a single neutrino flight and its periodicity is also resolvable for the concerned experiment, the potential will shift $\Delta m^2$ and $\theta$ by $E^{}_{\nu} \xi$ and $\xi E^{}_{\nu} / \Delta m^2$ respectively. No matter whether the ALP periodicity is resolvable for the experiment, one can always average over the observation time. In this case, the shifts of $\Delta m^2$ and $\theta$ are $(E^{}_{\nu} \xi)^2$ and $(\xi E^{}_{\nu} / \Delta m^2)^2$ respectively. If the ALP field oscillates very rapidly along the neutrino course but still remains in coherence, the oscillation probability will also be modified. 
The probability can either fluctuate with the amplitude $\xi/m^{}_{a}$ or averagely get distorted by the magnitude $(\xi/m^{}_{a})^2$.

\subsection{General Sensitivities}
In this section, 
 \begin{figure*}[t!]
 \centering
 \subfigure{\includegraphics[width=8cm]{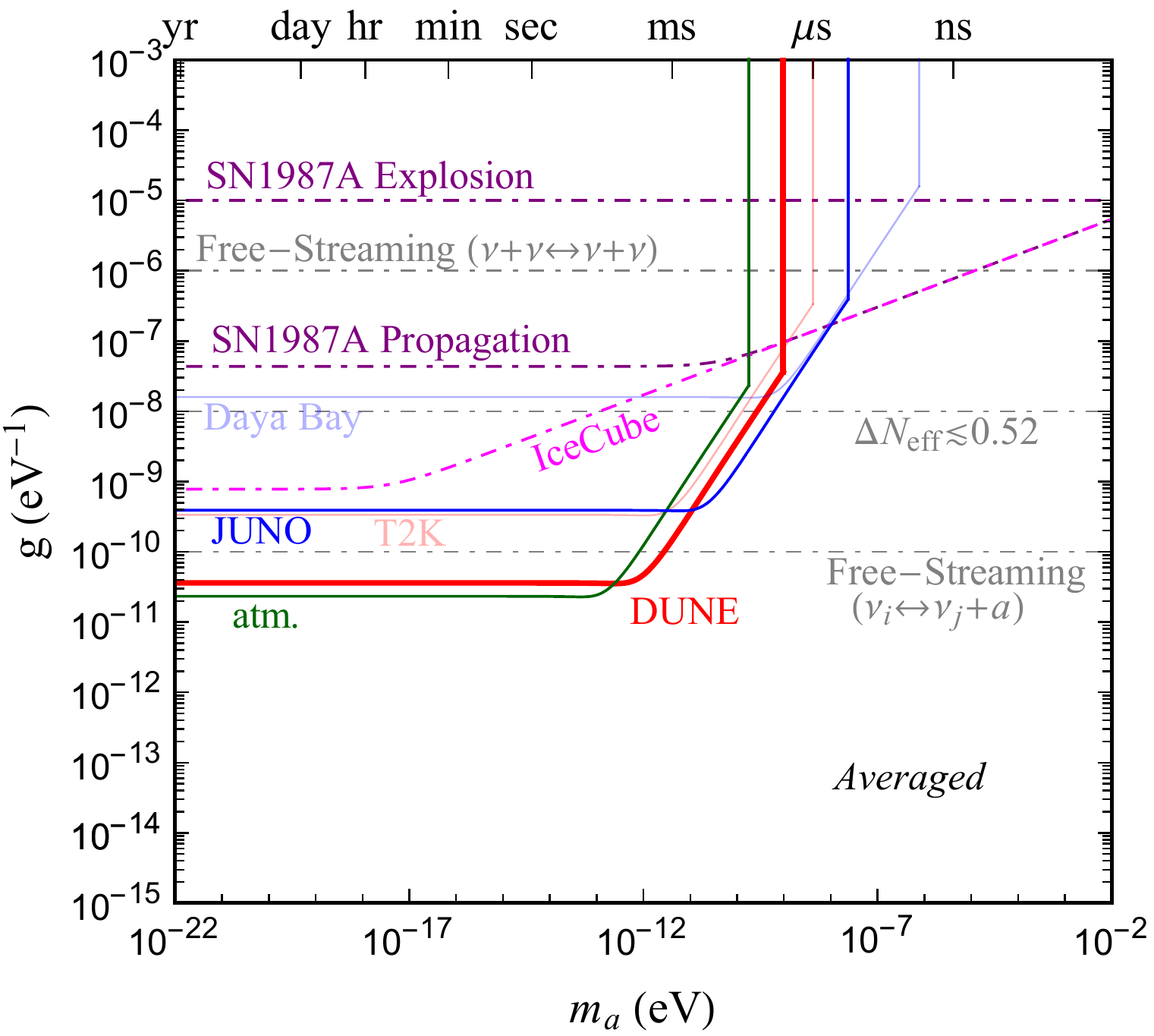}}
 \hspace{0.5cm}
 \subfigure{\includegraphics[width=8cm]{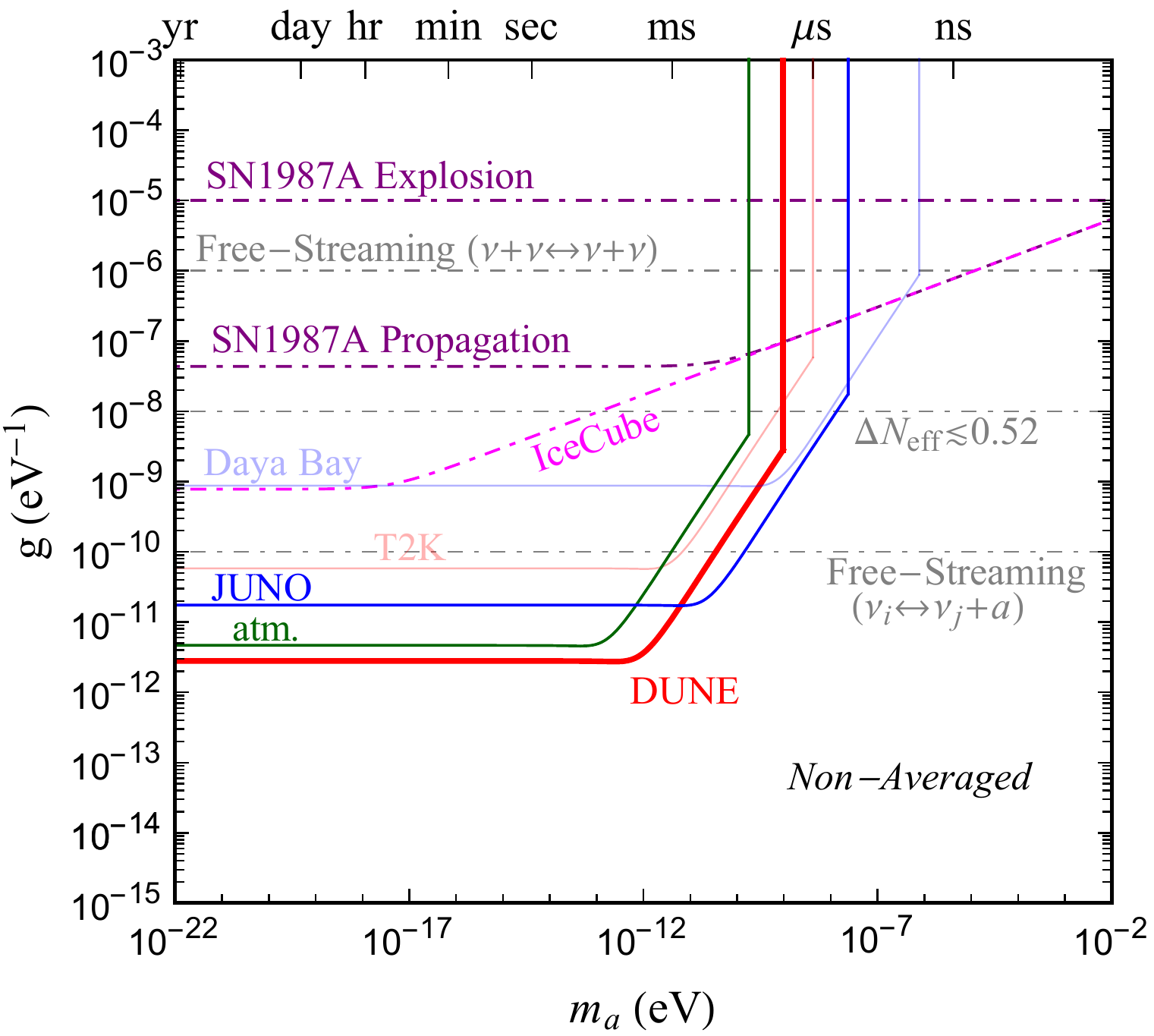}}
 \caption{\footnotesize The future sensitivities or constraints of the oscillation experiments as well as the astrophysical bounds. The \emph{left panel} stands for the case where the ALP-induced field is averaged over the observation time, while the \emph{right panel} is for the \emph{non-averaged} case. On the top axis, the corresponding periods of the ALPs with masses in the bottom axis have been marked. The red, blue and green curves represent the sensitivity of DUNE, JUNO, and the future atmospheric experiment respectively. In the left panel (right panel), the light red and light blue curves stand for the constraints (sensitivities) that can be made by T2K and Daya Bay, respectively. The dash-dotted curves are the astrophysical constraints including CMB (gray curves), IceCube-170922A (magenta curve), and SN1987A (purple curves). }
 \label{fig:GS}
 \end{figure*}
 we will roughly estimate the sensitivities (or constraints) of the oscillation experiments based on the analytical results of the last section. Suppose that an oscillation experiment can measure $\Delta m^2$ and $\theta$ with the $1\sigma$ experimental errors of $\sigma({\Delta m^2})$ and $\sigma({\theta})$. The experiment should be sensitive to the ALP coupling strength whose corrections to $\Delta m^2$ and $\theta$ are bigger than the experimental errors. To obtain the rough sensitivity, one can demand that the ALP-induced corrections should be smaller than errors of the relevant $\Delta m^2$ and $\theta$. This argument should be good enough as an order-of-magnitude estimate which is the main purpose of this section. Setting $\sigma\equiv {\rm min}\left[\sigma(\Delta m^2)/\Delta m^2, \sigma({\theta})\right]$, the argument can be translated into the following conclusive results for two different scenarios:
\begin{eqnarray} \label{eq:rqc1}
 g  \lesssim \frac{\Delta m^2}{2\pi E^{}_{\nu}} \frac{\sigma}{\sqrt{2\rho}} {\rm max}\left( 2\pi, m^{}_{a} t^{}_{\rm flt}\right) \hspace{2mm} \text{(\emph{Non-Averaged})} \hspace{1.5mm} 
\end{eqnarray}
when the ALP potential is either approximately constant or periodically resolvable; and
\begin{eqnarray} \label{eq:rqc2}
 g  \lesssim \frac{\Delta m^2}{2\pi E^{}_{\nu}} \sqrt{\frac{\sigma}{2\rho}} {\rm max}\left(2\pi, m^{}_{a} t^{}_{\rm flt}\right) \hspace{3mm} \text{(\emph{Averaged})} \hspace{3mm}
\end{eqnarray}
when the oscillating potential is averaged to the distortion effect. We have used a factor ${\rm max}\left( 2\pi, m^{}_{a} t^{}_{\rm flt}\right)$ and the condition $t^{}_{\rm flt} \approx {2\pi E^{}_{\nu}}/{\Delta m^2}$ to combine Cases I, II, III and Case IV, which is good enough as an order-of-magnitude estimate. With higher energy the oscillation experiment can have stronger sensitivity.
\begin{table}[b!]
\vspace{-5mm}
\begin{tabular}{|c|c|c|c|c|c|}
\hline
  & Daya Bay & T2K & JUNO & DUNE & atm. \\\hline
Baseline & $1.6~{\rm km}$& $295~{\rm km}$ & $53~{\rm km}$ & $1300~{\rm km}$ & $10$-$10^4~{\rm km}$ \\
\hline
Energy & $4~{\rm MeV}$ & $0.6~{\rm GeV}$ & $4~{\rm MeV}$ & $2.5~{\rm GeV}$ & $1$-$100~{\rm GeV}$ \\
\hline
$\sigma$ &  $\sim 0.3\%$ & $\sim 3\%$ & $\sim 0.2\%$ & $\sim 0.6\%$ & $\sim 4\%$ \\
\hline
\end{tabular}
\caption{\footnotesize Experimental Parameters}\label{tab:tab2}
\end{table}
One can observe that the sensitivity (constraint) can be made by an experiment depends on: (1) its baseline; (2) its beam energy; (3) the resolution power of the periodicity; (4) the statistics. We list in Table 1 for the necessary information of several representative experiments: Daya Bay \cite{An:2015rpe,An:2016ses}, T2K \cite{Abe:2018wpn}, JUNO \cite{Djurcic:2015vqa,An:2015jdp}, DUNE \cite{Acciarri:2015uup} and the future atmospheric experiments like ORCA \cite{Adrian-Martinez:2016fdl}, HK \cite{Abe:2018uyc} and PINGU \cite{Aartsen:2014oha}. For Daya Bay, the latest analysis \cite{An:2016ses} has yielded $\sin^2{2\theta^{}_{13}}=0.0841 \pm 0.0033$ and $\Delta m^2_{ee}=(2.5 \pm 0.085) \times 10^{-3}~{\rm eV^2}$, corresponding to $\sigma(\theta^{}_{13}) \approx 0.3\%$ and $\sigma(\Delta m^2_{ee})/\Delta m^2_{ee} \approx 3.4\%$. The T2K experiment \cite{Abe:2018wpn} has given their latest result $\sin^2{\theta^{}_{23}=0.526^{+0.032}_{-0.036}}$ and $\Delta m^2_{32} = (2.463 \pm 0.065) \times 10^{-3}~{\rm eV^2}$ corresponding to $\sigma(\theta^{}_{23}) \approx 3.4\%$ and $\sigma(\Delta m^2_{32})/\Delta m^2_{32} \approx 2.6\%$. For JUNO experiment, after 6 years of running the $1\sigma$ relative precisions of $0.54\%$, $0.24\%$ and $0.27\%$ can be achieved for $\sin^2{\theta^{}_{12}}$, $\Delta m^2_{21}$ and $\Delta m^2_{ee}$ \cite{An:2015jdp}, which correspond to $\sigma(\theta^{}_{12}) \approx 0.18\%$ and $\sigma(\Delta m^2_{21})/\Delta m^2_{21} \approx 0.24\%$. With an exposure of $150~{\rm kt\cdot MW\cdot yr}$ for DUNE, the resolutions of $\sin^2{2\theta^{}_{13}}$, $\sin^2{\theta^{}_{23}}$ and $\Delta m^{2}_{31}$ can reach $0.007$, $0.012$ and $1.6\times 10^{-5}~{\rm eV^2}$ \cite{Acciarri:2015uup}, respectively, corresponding to $\sigma(\theta^{}_{13}) \approx 0.63\%$, $\sigma(\theta^{}_{23}) \approx 1.2\%$ and $\sigma(\Delta m^2_{31})/\Delta m^2_{31} \approx 0.64\%$. For the atmospheric neutrino experiment, we will take its baseline and energy as $7\times 10^3~{\rm km}$ and $10~{\rm GeV}$ as an example. It is able to push the $1\sigma$ errors of $\sin^2{\theta^{}_{23}}$ and $\Delta m^{2}_{31}$ down to $0.05$ and $0.1\times 10^{-3}~{\rm eV^2}$ respectively \cite{Adrian-Martinez:2016fdl,Aartsen:2014oha}, corresponding to $\sigma(\theta^{}_{23}) \approx 5\%$ and $\sigma(\Delta m^2_{31})/\Delta m^2_{31} \approx 4\%$.

In Fig.~\ref{fig:GS}, the sensitivities or constraints of different oscillation experiments are demonstrated. For comparison, the existing astrophysical bounds which will be discussed in Section III have also been given. In the \emph{left panel}, the sensitivities and constraints are given based on Eq.~(\ref{eq:rqc2}) for the \emph{averaged} case, while the \emph{right panel} is based on Eq.~(\ref{eq:rqc1}) for the \emph{non-averaged} case. The red, blue and green curves demonstrate the future sensitivity of DUNE, JUNO, and the atmospheric neutrino experiment respectively. They can exceed the most stringent astrophysical bound given by the CMB free-streaming argument. One should keep in mind that the sensitivity study here is based on the rough arguments, while the exact one can only be done with the dedicated experimental simulation. 
%
\begin{figure*}[t!]
 \centering
 \includegraphics[width=16.8cm]{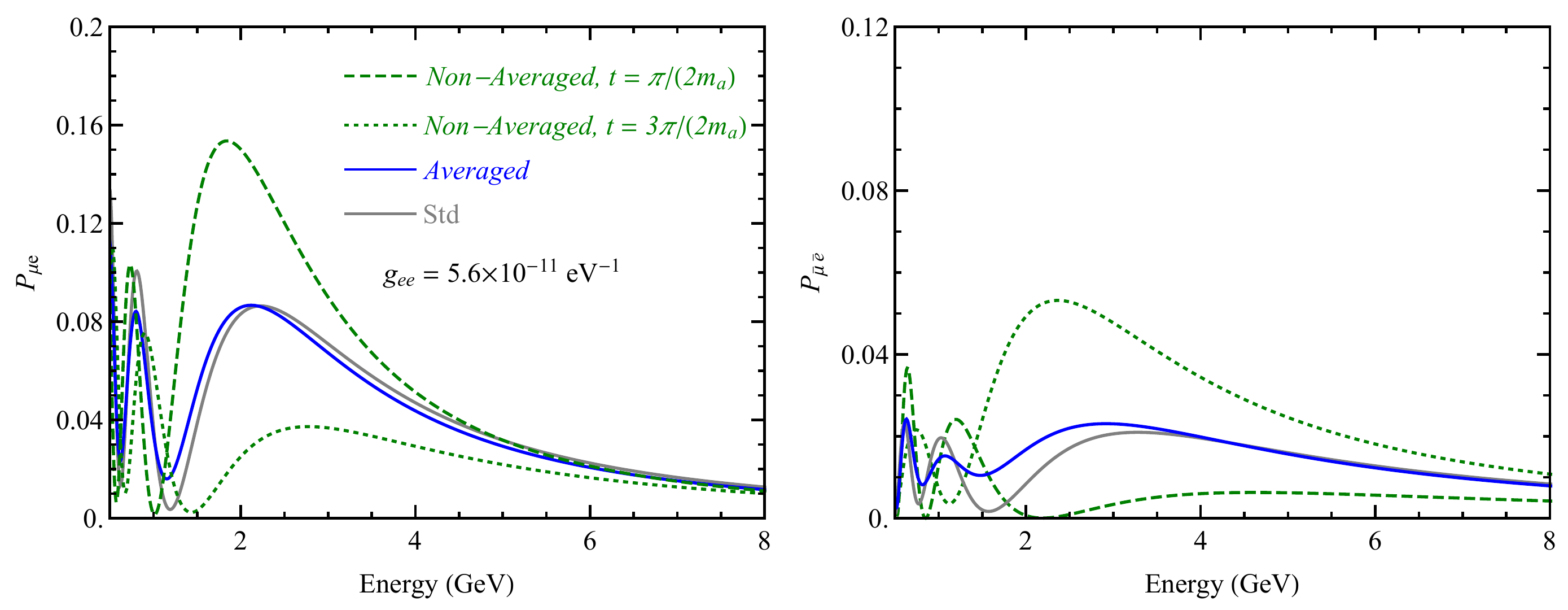}
\caption{\footnotesize  The oscillation probabilities of the appearance channels ${\nu}^{}_{\mu} \rightarrow \nu^{}_{e}$ (\emph{left panel}) and $\overline{\nu}^{}_{\mu} \rightarrow \overline{\nu}^{}_{e}$ (\emph{right panel}) for DUNE. The blue curves stand for the case where the ALP-induced field is \emph{averaged} over the observation time, whereas the green (dashed and dotted) curves signify the \emph{non-averaged} case. Also, the gray curves show the standard oscillation scenario without the ALP impact.}
\label{fig:ProbDune}
\end{figure*}
We will take DUNE as an example with detailed numerical simulations and analysis in the next section. We will find the rough estimation in this section is consistent in the order of magnitude with the result given by the detailed simulation.

\subsection{Impact on DUNE}
In this section, 
we will perform a detailed  study of the ALP-neutrino coupling parameters considering DUNE both at the probability as well as the $ \chi^{2} $ level. 
\begin{figure*}[t!]
\centering
\includegraphics[width=16.6cm]{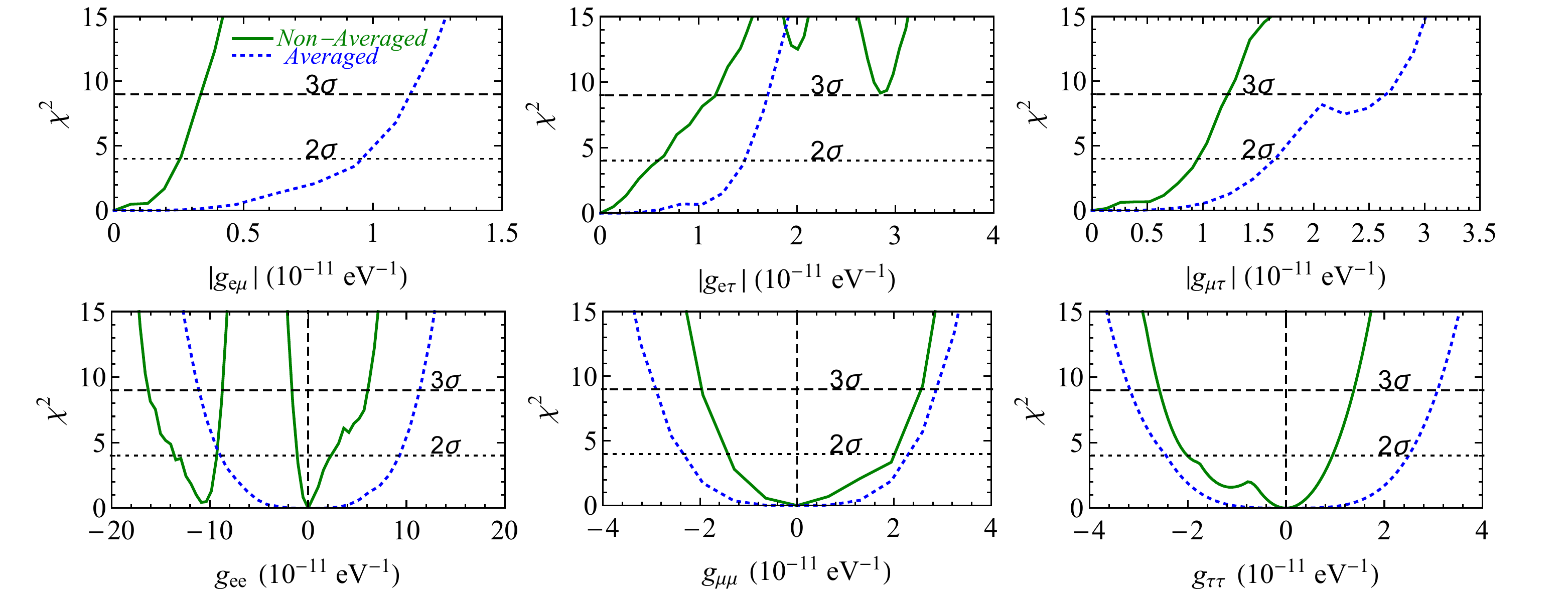}
\caption{\footnotesize  Sensitivities of ALP-neutrino coupling parameters $ g_{\alpha \beta} $ for DUNE. The dotted blue  curves stand for the case where the ALP-induced field is \emph{averaged} over the observation time, whereas the solid green curves signify the \emph{non-averaged} case.}
\label{fig:SensitivityDune}
\end{figure*}
DUNE is a proposed long baseline superbeam neutrino oscillation experiment at Fermilab \cite{Acciarri:2015uup, Alion:2016uaj}. It will have two neutrino detectors. The near detector will be placed near the source at Fermilab, while the far one will be installed 1300 km away from the neutrino source at Sanford Underground Research Facility (SURF) in Lead, South Dakota.
The far detector will utilize four $10$ kton liquid argon time projection chambers (LArTPCs).
Also, it will use the existing Neutrinos at the Main Injector (NuMI) beamline design at Fermilab as the neutrino source.

For the numerical analysis of DUNE, we use the
\texttt{GLoBES} package \cite{Huber:2004ka, Huber:2007ji} along 
with the required auxiliary files given in Ref.~\cite{Alion:2016uaj}.
Throughout this work we perform our simulation considering the 40 kton liquid argon detector. 
The flux corresponding to an 1.07 megawatt beam power gives
 $1.47\times 10^{21} $ protons on target (POT) per year due to an 80 GeV proton beam energy has been considered. 
The remaining experimental details like the signal and background normalization uncertainties for both the appearance and disappearance channels have been taken from DUNE CDR~\cite{Alion:2016uaj}. In addition, we consider a 3.5 years running time for each of the neutrino and antineutrino modes unless otherwise stated. 

\begin{table*}[t!]
\centering \scriptsize
\begin{tabular}{| c | c| c|}
\hline 
Parameters  &  \emph{Non-averaged}   & \emph{Averaged} \\
($ 10^{-11} $ eV$ ^{-1} $) &  2$ \sigma $\quad\quad\quad   3$ \sigma $  &  2$ \sigma $\quad\quad \quad  3$ \sigma $    \\
\hline  
$| g_{e \mu} |$  & $ < $ 0.25  \quad\quad $ < $ 0.34 & $ < $ 0.95  \quad\quad $ < $ 1.14\\
\hline 
$ |g_{e \tau} |$  & $ < $ 0.58 \quad\quad  $ < $ 1.16 & $ < $ 1.46 \quad\quad $ < $ 1.71\\
\hline 
$ |g_{\mu \tau} | $   &  $ < $ 0.95 \quad\quad  $ < $  1.23 &  $ < $  1.65 \quad \quad  $ < $ 2.65\\
\hline 
$  g_{ee} $  & (-13.62, -9.30)   \quad \quad    (-16.30, -8.71)     &   (-8.90, 9.30)\quad \quad   (-11.10, 11.30)\\ 
            & $ \oplus $(-1.11, 2.33) \quad\quad  $ \oplus $(-1.63, 6.15) & \\ 
\hline 
$ g_{\mu \mu} $  & (-1.42, 2.02)\quad\quad (-1.95, 2.56)& (-2.34, 2.34) \quad\quad (-2.90, 2.88) \\
\hline 
$ g_{\tau \tau} $  & (-2.01, 0.94)\quad \quad(-2.60, 1.37)& (-2.44, 2.50) \quad \quad(-3.18, 3.07) \\
\hline
\end{tabular}
\caption{\footnotesize The expected sensitivities of DUNE on the ALP-neutrino coupling terms $ g_{\alpha \beta} $ at the 2$ \sigma $ and the 3$ \sigma $ C.L. Here the second (third) column represents the sensitivities for the \emph{non-averaged} (\emph{averaged}) case. }
\label{tab:sens}
\end{table*}

To study the impact of the ALP-neutrino coupling on DUNE, we use the \texttt{GLoBES} extension file \textit{snu.c} as has been presented in Refs~\cite{Kopp:2006wp,Kopp:2007ne}. In our analysis, we have modified the NSI matter potential in \textit{snu.c} with 
the potential induced by the ALP field. We simulate the fake data of DUNE using the standard oscillation parameters $ \sin^2 \theta_{12} = 0.307,~\sin^2 \theta_{13} = 0.022,~\sin^2 \theta_{23} = 0.535,~\delta = - \pi/2,~\Delta m^{2}_{21} = 7.40 \times 10^{-5} ~{\rm eV^2}$ and $\Delta m^{2}_{31} = 2.50 \times 10^{-3} ~{\rm eV^2}$ which are compatible with the latest global-fit results~\cite{Capozzi:2016rtj,Esteban:2016qun,deSalas:2017kay}.
%
We marginalize the standard parameters $\theta_{23}$, $\delta$ and the mass hierarchy over their current 3$\sigma$ ranges in the test. As the remaining standard oscillation parameters have been measured with very high precisions, we keep them fixed in the analysis.
On the other hand, for the ALP-neutrino coupling parameters, we fix their true values as zero during the statistical analysis. For the fit we switch on only one of them at one time for numerical simplicity.
While performing the sensitivity study of the diagonal ALP-neutrino coupling parameters, we marginalize over the standard parameters in the fit.
 Moreover, for the off-diagonal parameters, we  also marginalize their phase in the range ($ 0 \rightarrow 2\pi $). All along this work, we perform our analysis considering two different scenarios for the ALP-induced potential, namely \emph{non-averaged} and \emph{averaged} cases. For the \emph{non-averaged} case, we take the ALP-induced potential $\xi^{}_{\alpha\beta}$ as a constant over time for simplicity. A more careful treatment would require the statistical analysis of the modulation on the time-binned data, which might be interesting for a future work. For the \emph{averaged} case, we have modified the \emph{snu.c} file by averaging the oscillation probability over the time points within one ALP cycle.

In Fig.~\ref{fig:ProbDune}, we show the oscillation probabilities of the appearance channels ($ {\nu_{\mu} \longrightarrow \nu_{e} } $ and $ {\overline{\nu}_{\mu} \longrightarrow \overline{\nu}_{e} } $) for DUNE. For illustration, we perform this study with a single non-zero representative value ($ g_{ee} = 5.6 \times 10^{-11} $ eV$ ^{-1} $) of the ALP-neutrino coupling term. The green (dashed and dotted) curves show the case where the ALP-induced potential is not averaged over the observation time (see the figure legend for more details). Whereas, the solid blue curves describe the scenario where the ALP-induced potential is averaged over time. Besides this, the standard case without the ALP-induced potential is shown as the gray solid curves for comparison.
For the \emph{non-averaged} case, one can observe significant deviations of the probabilities around the peak energy $\sim 2.5~{\rm GeV}$ compared to the standard case. When the ALP-induced potential oscillates as $g^{}_{ee} \sqrt{2\rho} \sin{m^{}_{a}t}$, the associated probability will vary roughly in between the two green curves. After averaging over the time, one would obtain the distortion effect as the blue curve in each panel, which is smaller than the \emph{non-averaged} case. 
As our intention at the probability level is for demonstration, hence we illustrate this analysis with a single non-zero ALP-neutrino coupling term for the appearance channel. 
Nevertheless, we will perform a detailed sensitivity study considering both the appearance and disappearance channels for all the coupling terms at the $ \chi^{2} $ level.

In Fig.~\ref{fig:SensitivityDune}, we show the sensitivity of DUNE to the six ALP-neutrino coupling parameters $ g_{\alpha \beta} $ in the $ \chi^{2} $-$ g_{\alpha \beta} $ plane. The top and bottom panels represent our results for the non-diagonal and the diagonal parameters, respectively. The green solid curves represent the \emph{non-averaged} case where a constant potential has been taken during the numerical simulation. The blue dotted curves signify the \emph{averaged} case where the ALP-induced potential is averaged over time in the final probability. Note that the black dotted and black dashed horizontal lines stand for the $ \chi^{2} $  values corresponding to the 2$ \sigma $ and the 3$ \sigma $ C.L., respectively. 
In Fig.~\ref{fig:SensitivityDune}, one can clearly observe that the \emph{averaged} case has looser sensitivity than the \emph{non-averaged} case. This can be easily understood from the analytical results before and from the probability demonstration in Fig.~\ref{fig:ProbDune}. It can be ascribed as the loss of information after averaging. 
From the first plot of the bottom row, we notice that for the green curve there is a local minimum around $ g_{ee} = -10.1 \times 10^{-11}$ eV$ ^{-1} $ other than a global minimum at zero. This shows a degenerate solution for $ g_{ee} $ and it arises from the unknown hierarchy (as we have marginalized over the mass hierarchy in the fit). In our careful analysis  by fixing the hierarchy to the normal one both in data and theory we find no degenerate solution for $ g_{ee} $. This tells that the lack of knowledge of the hierarchy may lead to a degenerate solution. Thus, for all these different cases we perform our numerical analysis over the marginalized hierarchy.
%
%
%
%
%
Also, comparing the top and the bottom row, we notice that overall sensitivity of the off-diagonal parameters is better than that of the diagonal ones. 
Table~\ref{tab:sens} summarizes the expected sensitivities for all the ALP-neutrino coupling parameters $ g_{\alpha \beta} $ for DUNE, where the 2$ \sigma $ and 3$ \sigma $ intervals are presented respectively. Note that the second and third panels of the table  show the sensitivity limit for the \emph{non-averaged} and \emph{averaged} cases, respectively.

\section{Astrophysical Bounds}
\subsection{Early Universe}
The cosmological evolution will be more or less modified, if neutrinos have strong interactions with an ultralight degree of freedom. During the expansion of the Universe, the thermal ALPs can be produced via the processes like $\nu+\nu \leftrightarrow a+a$ and $\nu_i \leftrightarrow \nu_j +a$ \footnote{We assume the neutrino mass is much larger than the ultralight ALP such that the process like ${\nu}+\nu \leftrightarrow a$ is kinematically forbidden.}, where $\nu_{i}$ and $\nu_{j}$ are the neutrino mass eigenstates with $m^{}_i > m^{}_j$. If the ALPs are fully thermalized before the neutrino decoupling around $1~{\rm MeV}$, they will contribute to the extra effective neutrino number by an amount of $\Delta N_{\rm eff} \approx 0.57$ \cite{Huang:2017egl}. The extra $N_{\rm eff}$ is strongly constrained by the observations of BBN and CMB. The current constraint of BBN is $\Delta N_{\rm eff} \lesssim 1$ at $95\%$ C.L. \cite{Mangano:2011ar}, while CMB can give a stronger constraint with the latest Planck 2018 result \cite{Aghanim:2018eyx}: $\Delta N_{\rm eff} \lesssim 0.52$ ($95\%$ C.L., $Planck$ TT+lowE). On the other hand, the strong interaction during the CMB epoch will also suppress the free-streaming length of neutrinos \cite{Hannestad:2005ex,Zhou:2011rc}, such that the evolution of CMB perturbations is severely disturbed. This can lead us to a very strong constraint on the coupling strength of the secret neutrino interactions between neutrinos and ALPs.

First, let us focus on the possible influence on $\Delta N_{\rm eff}$. We will find the binary process ${\nu}+\nu \leftrightarrow a+a$ is the dominant one for generating ALPs. The reaction rate of the binary process can be approximated as $\Gamma \approx g^4 m^{2}_{\nu} \cdot T^3$ with $T$ being the temperature of the neutrino plasma \footnote{The derivative coupling in Eq.~(\ref{eq:L}) is equivalent to the pseudoscalar coupling $h^{}_{\alpha\beta} a {\nu}^{}_{\alpha}\gamma^{}_{5}\nu^{}_{\beta}$ with a dimensionless coupling constant $h^{}_{\alpha\beta}=\sum^{}_{ij}U^{}_{\alpha i}U^{*}_{\beta j}(m^{}_{i}+m^{}_{j})g^{}_{\alpha\beta}$, only if each neutrino line in the Feynman diagram is attached to only one NG boson line \cite{Hannestad:2005ex,Farzan:2002wx}. However, this is not the case for the process ${\nu}+\nu \leftrightarrow a+a$. The reaction rate for the pseudoscalar coupling case is proportional to $T$ instead of $T^3$.}. During the epoch of radiation domination, the Hubble expansion rate is given by $H \approx 1.66 \sqrt{g^{}_*} T^2/M^{}_{\rm Pl}$, with $g^{}_*$ denoting the effective number of relativistic degrees of freedom at $T$ (refer to \cite{Steigman:2012nb,Laine:2006cp} for its values at various temperatures) and $M^{}_{\rm Pl} \simeq 1.221\times 10^{19}~{\rm GeV}$ being the Planck mass. Because of the ratio $\Gamma / H \propto T$, the thermal ALPs should first be copiously produced at high temperatures in the very beginning of the Universe. They should eventually freeze out at some low temperature $T^{}_{\rm fo}$ to evade the CMB constraint. The extra $N^{}_{\rm eff}$ contributed by the thermal ALPs is found to be 
 \begin{eqnarray}
\label{eq:Neff}
\Delta N^{}_{\rm eff} = \frac{1}{2} \times \frac{8}{7}\left[ \frac{g^{}_{*s}(T^{}_{\nu})}{g^{}_{*s}(T^{}_{\rm fo})} \right]^{4/3}
\end{eqnarray}
 for $T^{}_{\rm fo} > T^{}_{\nu}$, where $T^{}_{\nu}$ is the neutrino decoupling temperature around $1~{\rm MeV}$, and $g^{}_{*s} \approx g^{}_{*}$ holds before $T^{}_{\nu}$. To satisfy $\Delta N^{}_{\rm eff} \lesssim 0.52$, one can find the freeze-out temperature of thermal ALPs should be $T^{}_{\rm fo} \gtrsim 20~{\rm MeV}$ \cite{Zhang:2015wua}. Demanding $\Gamma \lesssim H$ at $20~{\rm MeV}$, we obtain the following constraint 
 \begin{eqnarray}
\label{eq:gbinary}
g  \lesssim 10^{-8} ~{\rm eV^{-1}}~\sqrt{\frac{0.05~{\rm eV}}{m^{}_{\nu}}}.
\end{eqnarray}
 The $\Delta N^{}_{\rm eff}$ generated by the freeze-in process $\nu_i \rightarrow \nu_j+a$ yields a negligible bound $g  \lesssim 10^{-2} ~{\rm eV^{-1}}$, therefore not elaborated here. 
 
 During the CMB era, the process ${\nu}+\nu \leftrightarrow a+a$ is suppressed by the low plasma temperature. There are mainly two processes that can reduce the neutrino free-streaming length: $\nu+\nu \leftrightarrow \nu + \nu$ and $\nu_i \leftrightarrow \nu_j +a$. For these two processes, the derivative coupling is equivalent to the pseudoscalar coupling. By requiring $\Gamma({{\nu}+\nu \leftrightarrow \nu + \nu}) \lesssim H$ at the photon decoupling temperature $T^{}_{\gamma} \approx 0.256~{\rm eV}$ \cite{Hannestad:2005ex,Ballesteros:2016xej}, one can obtain the constraint
\begin{eqnarray}
\label{eq:glesscmb1}
g \lesssim 10^{-6}~{\rm eV^{-1}}~\left(\frac{0.05~{\rm eV}}{m^{}_{\nu}} \right). \;
\end{eqnarray}
The decay process $\nu_i \leftrightarrow \nu_j +a$ is only relevant for the off-diagonal coupling $g^{}_{ij}$ with $i \neq j$ in the mass basis. The decay rate differs for different choices of neutrino mass patterns. If the three masses of neutrinos are very hierarchical, the decay rate reads $\Gamma^{}_{ij} \approx g^{2}_{ij} m^{4}_{i}/({48 \pi}T)$
with $g^{}_{ij} \equiv \sum^{}_{ij} U^{}_{\alpha i} U^{*}_{\beta j} g^{}_{\alpha\beta}$ and $m^{}_{i}$ being the mass of the heavier neutrino $\nu^{}_{i}$. To make sure the free propagation of neutrinos is sufficiently disrupted, one should include an angular factor $(m^{}_{i}/3T)^2$ \cite{Hannestad:2005ex}. Requiring $\Gamma_{\rm t} \equiv \Gamma^{}_{ij}(m^{}_{i}/3T)^2 \lesssim H$ at $T^{}_{\gamma}$, the following bound can be derived \cite{Hannestad:2005ex}:
\begin{eqnarray}
\label{eq:cmb2}
g^{}_{ij}\lesssim 10^{-10}~{\rm eV^{-1}}~\left(\frac{0.05~{\rm eV}}{m_{i}} \right)^3. \;
\end{eqnarray}
However, if the neutrino masses are degenerate, one should instead have 
\begin{eqnarray}
g^{}_{ij} &\lesssim & 10^{-10}~{\rm eV^{-1}}~\left(\frac{2.5 \times 10^{-3}~{\rm eV^2}}{\Delta m^{2}_{ij}} \right)^{3/2},\;
\end{eqnarray}
which is in the same order of magnitude as the the former one in Eq.~(\ref{eq:cmb2}). Overall, the most stringent constraint can be placed on the off-diagonal couplings in the mass basis $g^{}_{ij} \lesssim 10^{-10}~{\rm eV^{-1}}$ based on the free-streaming of CMB. The $\Delta N^{}_{\rm eff}$ limit of CMB can place bounds on all elements of $g^{}_{\alpha\beta}$ as $g  \lesssim 10^{-8} ~{\rm eV^{-1}}$. These early Universe bounds have been included in Fig.~\ref{fig:GS} as the dash-dotted gray curves.
\subsection{SN1987A Explosion}
The ALPs can also play an important role in the evolution of the supernovae. The constraints on the ultralight scalar coupling with neutrinos from the supernova neutrino observation SN1987A \cite{Hirata:1987hu,Bionta:1987qt} have been extensively studied in the literature, see Ref.~\cite{Farzan:2002wx,Zhou:2011rc} and the references therein for more details. The presence of the ALP-neutrino coupling can modify the standard evolution of supernova explosion by the following processes: (1) The ALPs can be thermally generated inside the core, carrying energy away from the core-collapse supernovae; (2) The ALPs can transport energy inside the supernova core and reduce the cooling time or the duration of the neutrino burst signal; (3) The spectrum of the observed neutrino flux might be distorted; (4) If the neutrino is the Majorana particle, the ALP-neutrino coupling can cause the the excessive deleptonization which will disable the supernova explosion. However, if the coupling is so strong that the ALPs is severely trapped inside the core, no constraint can be made on the contrary. The most stringent upper bound can be placed on the coupling $g^{}_{ee}$ from the energy loss argument as \cite{Farzan:2002wx} 
\begin{eqnarray}
\label{eq:gSN}
 g^{}_{ee} \lesssim  10^{-5}~{\rm eV^{-1}}~\left(\frac{0.05~{\rm eV}}{m^{}_{\nu}} \right), \;
\end{eqnarray}
which is much weaker than the limits of CMB. We have used Eq.~(\ref{eq:gSN}) to represent the constraining power of the SN1987A explosion in Fig.~\ref{fig:GS}.

\subsection{Galactic Propagation of Neutrinos from SN1987A and Blazar TXS 0506+056}
The propagation of astrophysical neutrino fluxes in the dark matter halo of our galaxy will be affected when the ALP-neutrino interaction is very strong \cite{Reynoso:2016hjr,Brdar:2017kbt,Arguelles:2017atb}. These neutrinos will suffer the energy loss and the direction change, due to the frequent scattering with the ALPs which forms the dark matter. The observed neutrino flux of SN1987A agrees well with the standard supernova neutrino theory, which should in turn give us a constraint on the ALP-neutrino coupling. On the other hand, the recent UHE neutrino event IceCube-170922A observed by IceCube coincides with a flaring blazar TXS 0506+056 with $\sim 3\sigma$ level \cite{IceCube:2018dnn}. The estimated neutrino luminosities are similar to that of the associated $\gamma$-rays, consistent with the prediction of blazar models \cite{Gaisser:1994yf,Achterberg:2006ik}. However, if there is a large attenuation effect for the UHE neutrino flux by scattering with ALPs, the original flux from TXS 0506+056 must be much stronger than the estimated one, which is inconsistent with the blazar observation. For the diffuse UHE neutrinos, the observed flux is around the Waxman-Bahcall (WB) bound \cite{Waxman:1998yy}. The large attenuation effect of ALP would require a diffuse flux to be much larger than the WB bound, which can in turn set a limit on the ALP-neutrino coupling. However, the original diffuse UHE neutrino flux might be well above the WB bound, since the complete sources of the UHE neutrino still remain unknown for us.  

A very simple argument would be that the mean free path of neutrinos $\lambda^{}_{\nu}$ should be smaller than the radius of the Milky Way $R^{}_{\rm MW}\sim 26~{\rm kpc}$. However, this is not true in general because the energy loss for each ALP-neutrino collision is unclear, which might be negligible compared with the initial neutrino energy in some case. For the scattering process ${\nu^{}_{i}}+a \rightarrow \nu^{}_{j} + a$, the average relative energy loss and angular change under one collision are estimated as
\begin{eqnarray}
\label{eq:scatterOneTime}
 \left<\frac{\Delta E^{}_{\nu}}{E^{}_{\nu}}\right> &\approx & \frac{m^{}_{a} E^{}_{\nu}}{s}+\frac{\Delta m^{2}_{ij}}{2s}, \\ \nonumber
 \left<\Delta \theta\right> &\approx & \frac{m^{}_{a}}{2\sqrt{s}}+\frac{\Delta m^{2}_{ij}}{4\sqrt{s}E^{}_{\nu}},\;
\end{eqnarray}
where $s\approx 2m^{}_{a}E^{}_{\nu}+m^{2}_{i}+m^{2}_{a}$ is the square of the center-of-mass energy. $\left<{\Delta E^{}_{\nu}}/{E^{}_{\nu}}\right>$ and $\left<\Delta \theta\right>$ are defined as the halves of their possible maximum values for the collision. One can easily find $\left<\Delta \theta\right> \ll \left<{\Delta E^{}_{\nu}}/{E^{}_{\nu}}\right>$ for our case, so we focus on the energy loss of neutrinos. Assuming $2m^{}_{a}E^{}_{\nu} \gtrsim m^{2}_{\nu}$ such that $s \approx 2m^{}_{a}E^{}_{\nu}$, we can have $\left<{\Delta E^{}_{\nu}}/{E^{}_{\nu}}\right> \approx \mathcal{O}(1)$. In this case, the energy loss of one single collision is significant, so one can adopt the argument $\lambda^{}_{\nu} \lesssim R^{}_{\rm MW}$. However, in the case of $2m^{}_{a}E^{}_{\nu} \ll m^{2}_{\nu}$ with $i = j$, one would have a negligible energy loss for each collision. It is difficult to transfer energy to a static ALP when the ALP mass is too small. Integrating over the energy loss along the neutrino course in our galaxy and in order not to violate the observations of SN1987A and blazar TXS 0506+056, a general requirement would be 
\begin{eqnarray}
\label{eq:blazarRq}
\lambda^{-1}_{\nu} R^{}_{\rm MW} \lesssim \left<\frac{E^{}_{\nu}}{\Delta E^{}_{\nu}} \right>,
\end{eqnarray}
where $\lambda^{}_{\nu} = \left(\sigma n^{}_{a} \right)^{-1}$. Here $\sigma \sim g^4 m^2_{\nu}$ is the approximated cross section for ${\nu}+a \leftrightarrow \nu + a$, which should be good enough for an order-of-magnitude estimation. The general bound can be formally written as
\begin{eqnarray}
\label{eq:gIC2}
 g &\lesssim & \left(\frac{m^{}_{a} \left<{E^{}_{\nu}}/{\Delta E^{}_{\nu}} \right>}{m^{2}_{\nu} \rho^{}_{\rm DM} R^{}_{\rm MW}  }\right)^{{1}/{4}}. \;
\end{eqnarray}
For an ALP mass $m^{}_{a} \ll m^2_{\nu}/(2E^{}_{\nu})$, one can obtain the following constraint
\begin{eqnarray}
\label{eq:gIC2}
 g &\lesssim & \left(\frac{1}{\rho^{}_{\rm DM} R^{}_{\rm MW} E^{}_{\nu}}\right)^{{1}/{4}} \\ \nonumber
 &\approx & 10^{-9}~{\rm eV^{-1}}~\left(\frac{290~{\rm TeV}}{E^{}_{\nu}}\right)^{{1}/{4}} \left(\frac{0.3~{\rm GeV \cdot cm^{-3}}}{\rho^{}_{\rm DM}}\right)^{{1}/{4}}. \;
\end{eqnarray}
Note that the constraint given by IceCube is almost two orders of magnitude stronger than that of the supernova under the same argument. This is simply due to the higher energy of the IceCube event ($\sim 290~{\rm TeV}$) compared with the supernova one ($\sim 30~{\rm MeV}$). The constraint with $m^{}_{\nu} \approx 0.05~{\rm eV}$ over the whole ALP mass range for IceCube can be found in Fig.~\ref{fig:GS} as the magenta curve. The corresponding supernova constraint along with the explosion one in the last section are shown as the purple curves in Fig.~\ref{fig:GS}.  

\section{Conclusion}
Assuming the derivative coupling between the ALP and the neutrinos, we have investigated the impacts of the ALP dark matter on neutrino oscillation experiments. Depending on the mass of the ALP, there are two different scenarios regarding the data analysis: the modulation effect induced by the oscillating ALP field can be resolved for the experiment (\emph{Non-Averaged}); the modulation effect is simply averaged to a distortion effect (\emph{Averaged}). Based on the simple argument, we find that the existing experiment like T2K can already exclude $g \gtrsim 3\times 10^{-10}~{\rm eV^{-1}}$ for $10^{-22}~{\rm eV} \lesssim m^{}_{a} \lesssim 10^{-11}~{\rm eV}$. The projected experiments like DUNE are sensitive to the coupling strength $g \gtrsim 10^{-12}~{\rm eV^{-1}}$ in the ALP mass range $10^{-22}~{\rm eV} \lesssim m^{}_{a} \lesssim 10^{-12}~{\rm eV}$ for the \emph{non-averaged} case. The $1\sigma$ sensitivity is reduced by a factor of $\mathcal{O}(10)$ for the \emph{averaged} case due to the cancellation of the oscillating ALP-induced potential. Using the \texttt{GLoBES} package, we have numerically simulated the data of DUNE. The sensitivity results on the six coupling parameters $g^{}_{\alpha\beta}$ for $\alpha,\beta=(e,\mu,\tau)$ have been yielded and summarized in Table~\ref{tab:sens}. The numerical results agree well in the order of magnitude with the simple estimation for DUNE. The impact of such coupling on the evolution of the early Universe has been discussed. A very stringent bound from the free-streaming of CMB can be made as $g^{}_{ij} \lesssim 10^{-10}~{\rm eV^{-1}}$. The propagation of neutrinos from SN1987A can put a constraint $g \lesssim 4\times 10^{-8}~{\rm eV^{-1}}$, while the IceCube observation can put a much stronger one $g \lesssim 7\times 10^{-10}~{\rm eV^{-1}}$. The next generation of neutrino experiments can probe the parameter range two orders of magnitude beyond the astrophysical limits.
%
\begin{acknowledgements}
The authors are indebted to Prof.~Shun Zhou for carefully reading this manuscript and for many valuable comments and suggestions. The authors thanks Prof.~Zhi-zhong Xing, Jing-yu Zhu and Xin Wang for insightful discussions.
GYH would like to thank Qin-rui Liu for helpful discussions. 
NN also thanks Dr. Sushant K. Raut and Dr. Mehedi Masud for useful discussion.
GYH is supported
by the National Natural Science Foundation of China under grant No. 11775232. The research work of NN was supported in part by the National Natural Science Foundation of China under Grant No. 11775231.
\end{acknowledgements}


\end{document}